\documentclass[twocolumn,amsmath,amssymb,superscriptaddress,prl,a4paper]{revtex4-1}
\usepackage{graphicx}
\usepackage[paperwidth=212mm,paperheight=297mm,centering,hmargin=1.65cm,vmargin=2.6cm]{geometry}

\usepackage{graphicx}                       
\usepackage{color}                          
\usepackage{dsfont}
\usepackage{mathrsfs}
\usepackage{bigints}
\usepackage{bbold}
\usepackage{amsthm}
\usepackage{setspace}

\newcommand{\ketbra}[2]{\left|#1\right\rangle\hskip-1mm\left\langle #2\right|}
\newcommand{\norm}[1]{\smash{:\hskip-1mm #1 \hskip-1mm :}}

\newcommand{\ket}[1]{\vert#1\rangle}
\newcommand{\bra}[1]{\langle#1\vert}

\newcommand{\id}{\mathds{1}}

\newcommand{\beq}{\begin{eqnarray}}
\newcommand{\eeq}{\end{eqnarray}}

\newcommand{\tr}{\operatorname{tr}}

\newcommand{\ps}{ps$^{-1}$}
\newcommand{\e}{\mathrm{e}}
\newcommand{\w}{\omega}

\begin{document}
\title{Vibrational enhancement of quadrature squeezing and phase sensitivity \\ in resonance fluorescence}

\author{Jake Iles-Smith}
\affiliation{Department of Physics and Astronomy, University of Sheffield, Sheffield, S3 7RH, United Kingdom}
\affiliation{{School of Physics and Astronomy, The University of Manchester, Oxford Road, Manchester M13 9PL, United Kingdom}}
\author{Ahsan Nazir}
\affiliation{{School of Physics and Astronomy, The University of Manchester, Oxford Road, Manchester M13 9PL, United Kingdom}}
\author{Dara P. S. McCutcheon}\email{dara.mccutcheon@bristol.ac.uk}
\affiliation{Quantum Engineering Technology Labs, H. H. Wills Physics Laboratory and Department of Electrical and Electronic Engineering, 
University of Bristol, BS8 1FD, United Kingdom}
\date{\today}
\pacs{Valid PACS appear here}

\begin{abstract}
\noindent{\bf{\sffamily{Abstract}}}\\
\noindent
Vibrational environments are commonly considered to be detrimental to the optical emission properties of 
solid-state and molecular systems, limiting their performance within quantum information protocols. 
Given that such environments arise naturally it is important to ask whether they can instead be turned to 
our advantage. Here we show that vibrational interactions can be harnessed within resonance fluorescence 
to generate optical states with a higher degree of quadrature squeezing than in isolated atomic systems. 
Considering the example of a driven quantum dot coupled to phonons, we demonstrate that it is feasible 
to surpass the maximum level of squeezing theoretically obtainable in an isolated atomic system and 
indeed come close to saturating the fundamental upper bound on squeezing from a two-level emitter. 
We analyse the performance of these vibrationally-enhanced squeezed states in a phase estimation 
protocol, finding that for the same photon flux, they can outperform the single mode squeezed vacuum state
\end{abstract}

\maketitle
\noindent{\bf{\sffamily{Introduction}}}\\
\noindent
Quadrature squeezed light has been identified as an important resource for continuous variable quantum information 
applications~\cite{PhysRevD.23.1693,Walls1983,RevModPhys.77.513,PhysRevLett.82.1784,RevModPhys.84.621,PhysRevLett.97.110501}. 
By reducing the variance of the electric field with respect to some phase below that of the vacuum, 
squeezed states can be used to increase accuracy in interferometric measurements for metrology applications~\cite{Goda2008}, 
for secret encodings in quantum key 
distribution~\cite{RevModPhys.84.621,PhysRevA.63.052311,PhysRevA.61.022309}, 
and are an essential resource in 
continuous variable quantum computing schemes~\cite{PhysRevLett.97.110501,PhysRevLett.82.1784}. 
Experimentally, squeezed states of light can be produced by a number of different methods~\cite{Andersen2016}, 
and a wide range of classes have been explored theoretically. 
These include the 
canonical Gaussian squeezed coherent states, 
as well as various non-Gaussian squeezed states obtained for example via photon addition or subtraction, 
or constructing superpositions of Gaussian coherent states~\cite{PhysRevA.78.063811}. 
To date, the record level of squeezing has been achieved in a squeezed vacuum state 
using an optical parametric amplifier~\cite{PhysRevLett.117.110801}. 
We note, however, 
that the level of squeezing is not the only consideration for applications~\cite{Lang2014}. 

In 1981 Walls and Zoller~\cite{Walls1981} investigated an intriguing source of non-Gaussian 
quadrature squeezing, 
which is generated in the multimode resonance fluorescence field of a driven two-level emitter (TLE), 
where the emitted photons are antibunched~\cite{PhysRev.188.1969,Dalibard1983}. 
This was recently demonstrated experimentally using a
semiconductor quantum dot platform~\cite{Schulte2015}, which offers the necessary high photon collection efficiency 
unobtainable in most atomic approaches~\cite{lodahl2015interfacing,PhysRevLett.108.093602,Matthiesen2013}.  
The TLE scheme relies on the build-up of steady-state coherence between 
the ground and excited state, 
i.e. a state of the form $\ket{g}+c\,\ket{e}$ with some appreciable $c$~\cite{Walls1981}. 
This coherence, as well as the saturability of the emitter (which leads to antibunching) 
are inherited by the field, and together these properties give rise 
to photon statistics which amount to squeezing of a particular field quadrature. 
In the standard regime of atomic resonance fluorescence 
only a restricted set of atomic coherences (values of $c$) can be explored, and  
squeezing is achievable only in the weak-driving Rayleigh (equivalently Heitler) 
limit~\cite{Matthiesen2013,PhysRevLett.108.093602,Konthasinghe2012,McCutcheon2013,iles2016fundamental,nazir2015modelling,Bennett2016,PhysRevLett.114.067401}, 
with squeezing values considerably smaller than the 
fundamental bound for a two-level system~\cite{Collett1984,Ficek:84,Aravind:87,Wadkiewicz1987,PhysRevA.88.023837,PhysRevLett.109.013601}. 
Furthermore, it has yet to be explored what applications might make use of states produced 
in this way, and whether they offer any advantages over the more commonly studied single mode 
squeezed coherent states. 

In this work we establish how to generate  
quadrature squeezed states 
at stronger driving above saturation, 
resulting in levels of squeezing that can surpass the (atomic) Walls and Zoller maximum. 
Our approach harnesses the vibrational environment commonly present in solid-state 
and molecular emitters to access states otherwise unreachable in conventional resonance fluorescence. 
In particular, we exploit thermalisation processes within the driving-induced dressed state basis 
to obtain a non-equilibrium steady-state with significant coherence above saturation~\cite{McCutcheon2013}, 
where it would usually be strongly suppressed. 
Moreover, we analyse the performance of these resonance fluorescence states in a
simple phase estimation protocol, finding that they are able to outperform the single mode squeezed vacuum state, 
and that this is only achievable with the vibrational interactions included. 
As a concrete example, 
we illustrate this behaviour through a microscopic model of a driven semiconductor quantum dot coupled to a phonon 
environment~\cite{PhysRevLett.105.177402,PhysRevLett.104.017402,1367-2630-12-11-113042,Roy2011X}. 
We show that off-resonant driving can be used to access levels of squeezing 
close to the fundamental upper bound for a two-level system, beyond those possible in the 
Rayleigh scattering limit below saturation (and hence any parameters  
for which the vibrational environment is absent). 
We thus identify a scenario in which vibrational processes in solid-state emitters can be 
used to generate squeezed states and enhance phase sensitivity in a way not possible 
in their absence.

\noindent{\bf{\sffamily{Results}}}\\
\noindent
{\small{{\bf{Squeezing in atomic resonance fluorescence.}}}} 
Let us begin by introducing the basics of squeezing, and how it can arise in the field produced by 
a TLE in the standard setting without any additional vibrational environment~\cite{Walls1981}. 
The electric field quadrature relative to a phase reference $\varphi$ is defined as  
$E(t,\varphi) = \e^{i \varphi}E^{(+)}(t)+\e^{-i \varphi}E^{(-)}(t)$ with $E^{(+)}(t)$ the positive frequency component of the electric field, 
and the quadrature variance is $\Delta E(t,\varphi)^2=\langle E(t,\varphi)^2\rangle -\langle E(t,\varphi)\rangle^2$. 
The Heisenberg uncertainty relation bounds the variances in 
two out of phase quadratures via 
$\Delta E(t,\varphi)\Delta E(t,\varphi+\pi/2)\geq\frac{1}{2}|\langle [E(t,\varphi),E(t,\varphi+\pi/2)]\rangle |$. 
We say that a state of the field is a minimal uncertainty state when the 
bound is saturated, 
and the field is said to be squeezed if there is a quadrature that satisfies 
$\Delta E(t,\varphi)^2<\frac{1}{2}|\langle [E(t,\varphi),E(t,\varphi+\pi/2)]\rangle |$. 
When considering emission from a TLE 
within the dipole approximation and the far field limit, the field 
may be written in the Heisenberg picture 
${E}^{(+)}(t) =E_0(t)-\sqrt{2\Gamma/\pi}\sigma(t)$~\cite{Kiraz2004,walls2008quantum}. 
The first term describes the field in the absence of the TLE, which we assume to be in the vacuum state. 
The second term describes the TLE emission, where the dipole operator is $\sigma = \ket{g}\bra{e}$ 
with the ground state $\ket{g}$ and excited state $\ket{e}$, 
and $\Gamma$ is the spontaneous emission rate. 
The uncertainty relation for the field in the steady state can be written in terms of the TLE quadrature, $X(\varphi)=\e^{i\varphi}\sigma+\e^{-i\varphi}\sigma^{\dagger}$, as
\beq
\Delta X(\varphi)\Delta X(\varphi+\pi/2)\geq |\langle 2\sigma^{\dagger}\sigma-\openone \rangle |,
\label{Heisenberg}
\eeq
where expectation values are taken in the long time limit. 
For the TLE quadrature we have $\Delta X(\varphi)^2=1-\langle X(\varphi) \rangle^2$, 
showing that the quadrature with the smallest fluctuations is that with the greatest expectation value, 
meaning that any squeezing can therefore be thought of as amplitude squeezing. 
Without loss of generality we can write $\langle \sigma \rangle =|\langle \sigma \rangle|\e^{-i \phi}$, 
from which we find $\langle X(\varphi)\rangle^2=4|\langle \sigma \rangle|^2 \cos^2(\phi-\varphi)$, and 
we see that the 
quadratures with the lowest and largest variances have 
$\langle X(\varphi=\phi)\rangle ^2=4|\langle \sigma \rangle|^2$ and 
$\langle X(\varphi=\phi+\pi/2)\rangle^2=0$ respectively. 
The squeezing condition becomes equivalent  
to $\norm{\Delta X(\phi)^2}<0$, where we define the normally ordered quantity 
$\norm{\Delta X(\phi)^2}=\Delta X(\phi)^2-|\langle 2\sigma^{\dagger}\sigma-1 \rangle|$. 
The angle $\phi$ that defines the minimum uncertainty quadrature is the 
phase of the TLE dipole, which depends on the detuning of the driving laser from the transition energy, and 
is discussed further below.  

Before we determine the conditions under which squeezing occurs, it is instructive to 
establish a relationship between the quadrature variance and quantities more commonly used in studies of 
resonance fluorescence: namely the power, the coherent scattering, and antibunching. 
To do so we recall that the steady-state spectrum of 
resonance fluorescence from a TLE is proportional to 
$I(\w)=\frac{1}{\pi}\mathrm{Re}\int_0^{\infty}\mathrm{d}\tau g^{(1)}(\tau)\e^{-i\w\tau}$, 
with $g^{(1)}(\tau)=\langle\sigma^{\dagger}(\tau)\sigma\rangle$ the first order field correlation function. 
The coherent contribution is separated by writing 
$I(\w)=I_{\mathrm{coh}}(\w)+I_{\mathrm{inc}}(\w)$ where 
$I_{\mathrm{coh}}(\w)=g^{(1)}_{\mathrm{coh}} \delta(\w)$ and 
$I_{\mathrm{inc}}(\w)=\frac{1}{\pi}\mathrm{Re}\int_0^{\infty}\mathrm{d}\tau [g^{(1)}(\tau)-g^{(1)}_{\mathrm{coh}}]\e^{-i\w\tau}$, with 
$g^{(1)}_{\mathrm{coh}}=\lim_{\tau\to\infty}g^{(1)}(\tau)=|\langle \sigma \rangle |^2$~\cite{Konthasinghe2012,McCutcheon2013,iles2016fundamental,nazir2015modelling}. 
The total radiated power can be similarly separated into coherent and incoherent contributions, giving 
$P=\int_{-\infty}^{\infty}\mathrm{d}\w I(\w)=\langle \sigma^{\dagger}\sigma\rangle = P_{\mathrm{coh}}+P_{\mathrm{inc}}$ with 
$P_{\mathrm{coh}}=|\langle \sigma \rangle |^2\leq P$. 
We note that defined in this way, these are dimensionless power contributions satisfying $0\leq P \leq 1$ and $0\leq P_{\mathrm{coh}} \leq 0.25$. 
In terms of these power contributions the 
quadrature with the minimum variance has 
\beq
\norm{\Delta X(\phi)^2}=1-|2P-1|-4P_{\mathrm{coh}}, 
\label{NormXVar}
\eeq
showing that squeezing only occurs when $P_{\mathrm{coh}}$ is appreciable, 
and the total power $P$ takes 
on values chose to $0$ or $1$. 
Finally, we note that the antibunching behaviour 
is captured by the probability to simultaneously detect two photons $g^{(2)}(0)=\langle (\sigma^{\dagger})^2 \sigma^2\rangle$. 
For a TLE $\sigma^2=0$, leading to $g^{(2)}(0)=0$ regardless of the parameter regime. 

For a TLE without additional vibrational interactions, 
the effectively flat frequency spectrum of the electromagnetic environment 
restricts the magnitude of the total and coherently scattered power,  
which in turn limits the level of squeezing. 
To see this, we consider a TLE 
described by a density operator $\rho$ 
driven by a coherent source 
with Rabi frequency $\Omega$ and laser--TLE detuning $\delta$, for which a 
zero temperature 
master equation can be written within a rotating frame and the rotating wave approximation as
$
\dot\rho = -(i/\hbar)\left[H_\mathrm{S},\rho\right] +\Gamma\mathcal{L}_\sigma[\rho].
$
Here $H_\mathrm{S}=\hbar\delta\sigma^\dagger\sigma + \frac{\hbar}{2}(\sigma \Omega+\sigma^{\dagger}\Omega^*)$, 
and the emission dissipator is $\mathcal{L}_\sigma[\rho] =\sigma\rho\sigma^{\dagger} - (1/2)(\sigma^\dagger\sigma\rho+\rho\sigma^\dagger\sigma)$, 
whose driving and detuning independent form is a consequence of the flat spectrum assumption. 
Solving the master equation in the steady state, 
i.e. when $\dot\rho = 0$, we find $P=\mathcal{S}/(2(\mathcal{S}+1))$ and $P_{\mathrm{coh}}=P/(\mathcal{S}+1)$, leading to
\beq
\norm{\Delta X(\phi)^2} = \frac{\mathcal{S}(\mathcal{S}-1)}{(\mathcal{S}+1)^2},
\label{DeltaXWalls}
\eeq
where we have defined the saturation parameter $\mathcal{S}=s/(1+d)$ in terms of a 
dimensionless driving $s=2(\Omega/\Gamma)^2$ and detuning $d=4(\delta/\Gamma)^2$. 
The squeezing is greatest when Eq.~({\ref{DeltaXWalls}}) is minimised, which occurs for $\mathcal{S}=1/3$, at which point 
$P=1/8$, $P_{\mathrm{coh}}=3/32$, and 
$\norm{\Delta X(\phi)^2}=-0.125$~\cite{Walls1981}. 
This represents the theoretical maximum squeezing obtainable in resonance fluorescence for a TLE undergoing 
spontaneous emission into an unstructured environment~\cite{Walls1981}. 
Furthermore, the Heisenberg uncertainty relation in Eq.~({\ref{Heisenberg}}) 
in this simple case becomes $\sqrt{\mathcal{S}^2+1}\geq 1$, which is saturated only 
when the TLE is undriven and $\mathcal{S}=0$, and we therefore conclude that 
although the emitted field is squeezed when $\mathcal{S}=1/3$ ($\sqrt{\mathcal{S}^2+1}\approx 1.05$), it is close to, but 
not in a minimal uncertainty state. 
\\\\
\noindent
{\small{{\bf{Vibrational enhancement of squeezing.}}}} 
As has been previously noted~\cite{Collett1984,Ficek:84,Aravind:87,Wadkiewicz1987,PhysRevA.88.023837,PhysRevLett.109.013601}, 
the squeezed state obtained for a simple TLE as described above is not optimal, and 
is a consequence of the limited set of states available 
in this simple model. To verify this, we consider a TLE described by a completely generic density operator, 
i.e.~one that is not necessarily a solution to the simple master equation above. 
Explicitly, we write
\begin{align}
\rho=\frac{1}{2}\Big(\openone+l \big[\cos\theta(2\sigma^{\dagger}\sigma-\openone) 
+ \sin\theta X(\phi) \big]\Big),
\label{genrho}
\end{align}
parameterised by a Bloch vector length $0\leq l\leq1$, polar angle $0\leq\theta\leq\pi$ and dipole phase $0\leq\phi\leq 2\pi$. 
We can then express the total power as 
$P=\langle\sigma^{\dagger}\sigma\rangle=(1/2)(1+ l \cos\theta)$ and the coherently scattered power 
as $P_{\mathrm{coh}}=|\langle\sigma\rangle|^2=(1/4)l^2 \sin^2\theta$. 
These expressions represent a less restricted set of 
values for the power contributions than before, and 
minimising Eq.~({\ref{NormXVar}}) now gives 
$l=1$, and $\theta= \pi/3$ ($P_{\mathrm{coh}}=3/16$, $P=3/4$) or 
$\theta=2\pi/3$ ($P_{\mathrm{coh}}=3/16$, $P=1/4$), both of which result in 
$\norm{\Delta X(\phi)^2}=-0.25$. 
This is the true maximum level of squeezing obtainable 
from a TLE, and is limited only by the two level nature of the system~\cite{Collett1984,Ficek:84,Aravind:87,Wadkiewicz1987}. 
Moreover, the uncertainty relation of Eq.~({\ref{Heisenberg}}) now reduces to 
$\smash{\sqrt{1-l^2\sin^2\theta}}\geq | l \cos\theta|$. 
This is saturated for all 
$\theta$ when $l=1$, for which $\rho$ describes a pure state, 
verifying that the states with maximum squeezing have minimum uncertainty. 
More generally we see that the Bloch vector length $l$ can be used to parameterise 
how close a state is to minimum uncertainty. 

How, then, can we obtain such a state within resonance fluorescence? 
We shall now show that naturally occurring vibrational interactions in, 
for example, solid-state and molecular systems can 
help to drive a TLE into such a state, resulting in a level of 
squeezing close to the maximum value of $\norm{\Delta X(\phi)^2}=-0.25$, 
and certainly greater than the value of $\norm{\Delta X(\phi)^2}=-0.125$ obtainable 
in their absence. This can be understood qualitatively by considering equilibration of our system 
with respect to the additional vibrational environment. 
The density operator for a TLE driven according to the Hamiltonian $H_\mathrm{S}$ defined previously, 
but now reaching thermal equilibrium due to some additional reservoir, can be written as the thermal state 
\beq
\rho_{\mathrm{th}}=\frac{\e^{-\beta H_\mathrm{S}}}{\mathrm{Tr}(\e^{-\beta H_\mathrm{S}})},
\label{rhoth}
\eeq
where $\beta = 1/k_\mathrm{B} T$ is the inverse temperature. 
After performing the necessary exponentiation, we see that $\rho_{\mathrm{th}}$ can be written 
in the general form shown in Eq.~({\ref{genrho}}), with the Bloch vector 
parameters given by $l=l_{\mathrm{th}}=\tanh(\hbar\beta\eta/2)$ with $\eta=\sqrt{\delta^2+\Omega^2}$, 
$\theta = \arctan(\Omega/\delta)$ and 
$\phi = \mathrm{arg}(\Omega)$. 
As such, we see that 
for suitable choices of the adjustable Hamiltonian parameters 
$\Omega$ and $\delta$, 
we can satisfy the conditions above which lead 
to maximum squeezing. Explicitly, we find 
\beq
\norm{\Delta X(\phi)^2}=1-\frac{|\delta|}{\eta} l_{\mathrm{th}} -\frac{|\Omega|^2}{\eta^2}l_{\mathrm{th}}^2,
\label{DeltaXApprox}
\eeq
and if we choose $\delta=\pm\Omega/\sqrt{3}$, 
a minimal uncertainty state with the maximum squeezing of 
$\norm{\Delta X(\phi)^2}=-0.25$ is achieved in the low temperature limit ($\hbar\beta\eta\to \infty$, $l_{\mathrm{th}}\to 1$), 
while the quadrature displaying this squeezing can be chosen by adjusting the phase of $\Omega$. 
We emphasise that these states are achieved in the steady-state, 
and as such persist for as long as the driving and temperature conditions stay fixed. 
\\\\
\noindent
{\small{{\bf{Squeezing from a driven quantum dot.}}}}
We now explore the vibrational enhancement of squeezing in greater detail, 
and analyse the limitations of the simple argument given  
above. To do so 
we consider the example of a semiconductor quantum dot (QD) 
as a solid-state TLE with ground state $\ket{g}$  
and excited state $\ket{e}$ 
describing a single exciton of energy $\hbar\omega_0$.
The QD is driven by a semiclassical monochromatic laser. 
Within a frame rotating with respect to the laser frequency and after applying the rotating wave approximation   
the system Hamiltonian may be written as 
$H_\mathrm{S} =\hbar\delta\sigma^\dagger\sigma+\frac{\hbar}{2}(\sigma\Omega+\sigma^{\dagger}\Omega^*)$, as before. 
The QD couples to both vibrational and electromagnetic environments, 
with each steering the system towards competing equilibrium states. 
We make use of a variational master equation technique which allows regimes 
of strong QD--phonon coupling and laser driving to be explored within a 
robust formalism~\cite{PhysRevB.84.081305,McCutcheon2013,nazir2015modelling}. 
Full details of our model can be found in Methods and Supplementary Notes 1 and 2. 
The result is a master equation describing the QD excitonic degrees of freedom 
of the form 
$
\dot\rho_\mathrm{V}= -(i/\hbar)\left[H_\mathrm{r},\rho_\mathrm{V}\right] 
+ \mathcal{K}_{\mathrm{ph}}[\rho_\mathrm{V}] + {\Gamma}\mathcal{L}_\sigma[\rho_\mathrm{V}],
$
where $\rho_\mathrm{V}$ is the reduced density operator of the QD in the variational frame and 
$H_\mathrm{r}=\hbar\delta_\mathrm{r}\sigma^{\dagger}\sigma+\frac{\hbar}{2}(\sigma\Omega_\mathrm{r}+\sigma^{\dagger}\Omega_\mathrm{r}^*)$. 
Coupling to phonons is 
accounted for in the renormalised 
detuning $\delta_\mathrm{r}$ and driving $\Omega_\mathrm{r}$, together with  
the phonon dissipator $\mathcal{K}_{\mathrm{ph}}$. 
The form of this dissipator is given in Methods, and includes various phonon 
assisted relaxation and dephasing processes. The stregnths of these processes depend on the phonon spectral density, 
which is a measure of the electron-phonon coupling strength weighted by the phonon density of states, and 
takes the form $J(\w)\propto \w^3 \exp[-(\w/\w_\mathrm{c})^2]$ with cut-off frequency 
$\omega_\mathrm{c}$ that is inversely proportional to the exciton size~\cite{nazir2015modelling}, 
and which sets the timescale of lattice relaxation. 
In the long-time limit the effect of the phonon dissipator is to tend the QD exciton 
towards thermal equilibrium with respect 
to the dressed eigenstates of $H_\mathrm{r}$, while the spontaneous emission term  
${\Gamma}\mathcal{L}_\sigma[\rho_\mathrm{V}]$ acts to move the system towards 
the ground state. 

\begin{figure}
	\includegraphics[width=0.95\columnwidth]{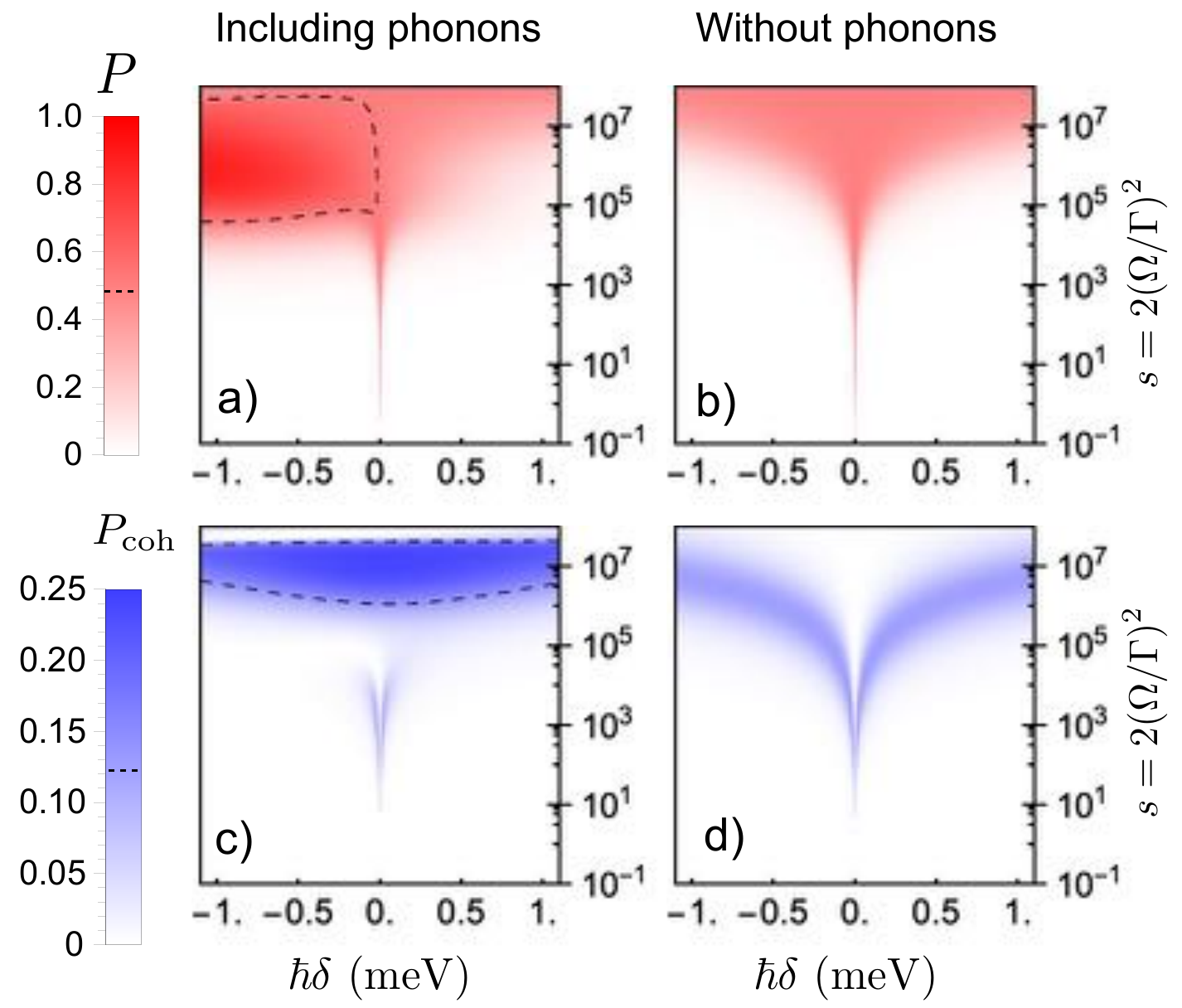}
	\caption{{\bf{Resonance fluorescence power contributions.}} 
	Normalised total emission power $P$ with a) and without b) phonons, and 
	coherently scattered power $P_{\mathrm{coh}}$ with c) and without d) phonons, plotted   
	as functions of the scaled driving strength [$s=2(\Omega/\Gamma)^2$] and detuning. 
	Comparing the cases with and without phonons we see markedly behaviour at large driving strengths.
	The dashed contour lines 
	indicate half the maximum possible value for each power contribution from a two-level-emitter, 
	$\frac{1}{2}P^{\mathrm{max}}=0.5$ and $\frac{1}{2}P_{\mathrm{coh}}^{\mathrm{max}}=0.125$, 
	neither of which are exceeded in the absence of phonons. 
	Parameters: $\alpha = 0.027$~ps$^2$, $\omega_\mathrm{c} = 2.2$~\ps,  $T=4$~K, and $1/\Gamma =700$~ps. 
	}
	\label{qd_squeeze}
\end{figure}

Having outlined our model, in Fig.~\ref{qd_squeeze} we show the dimensionless total power 
$P$ and the coherently scattered power $P_{\mathrm{coh}}$ as functions of the dimensionless driving $s$ 
and the detuning $\delta$, 
both including [a) and c)] and excluding [b) and d)] phonons as 
indicated. In all cases with phonons included we define the detuning relative to a polaron shift, 
i.e.~$\delta\rightarrow\delta-\int_0^{\infty}d\omega J(\omega)/\omega$. 
As noted for resonant driving in Ref.~\cite{McCutcheon2013}, plot 
c) shows that in the presence of phonons we obtain a significant coherent scattering power 
when driving above saturation, which we here find extends across a broad range of 
detunings as well. 
As expected, this occurs as a result of the phonons 
attempting to thermalise the QD exciton 
in the dressed state basis, which at low temperatures and strong fields 
leads to sustained steady-state coherence. At very high driving strengths the coherent scattering falls off, 
as here the generalised Rabi frequency $\sqrt{\Omega_\mathrm{r}^2+\delta_\mathrm{r}^2}$ exceeds the 
extent of the phonon spectral density set by the cut-off $\w_\mathrm{c}$, leading to a regime in which the exciton and 
phonons are effectively decoupled~\cite{vagov2007,PhysRevB.84.081305}.

\begin{figure}
	\includegraphics[width=0.95\columnwidth]{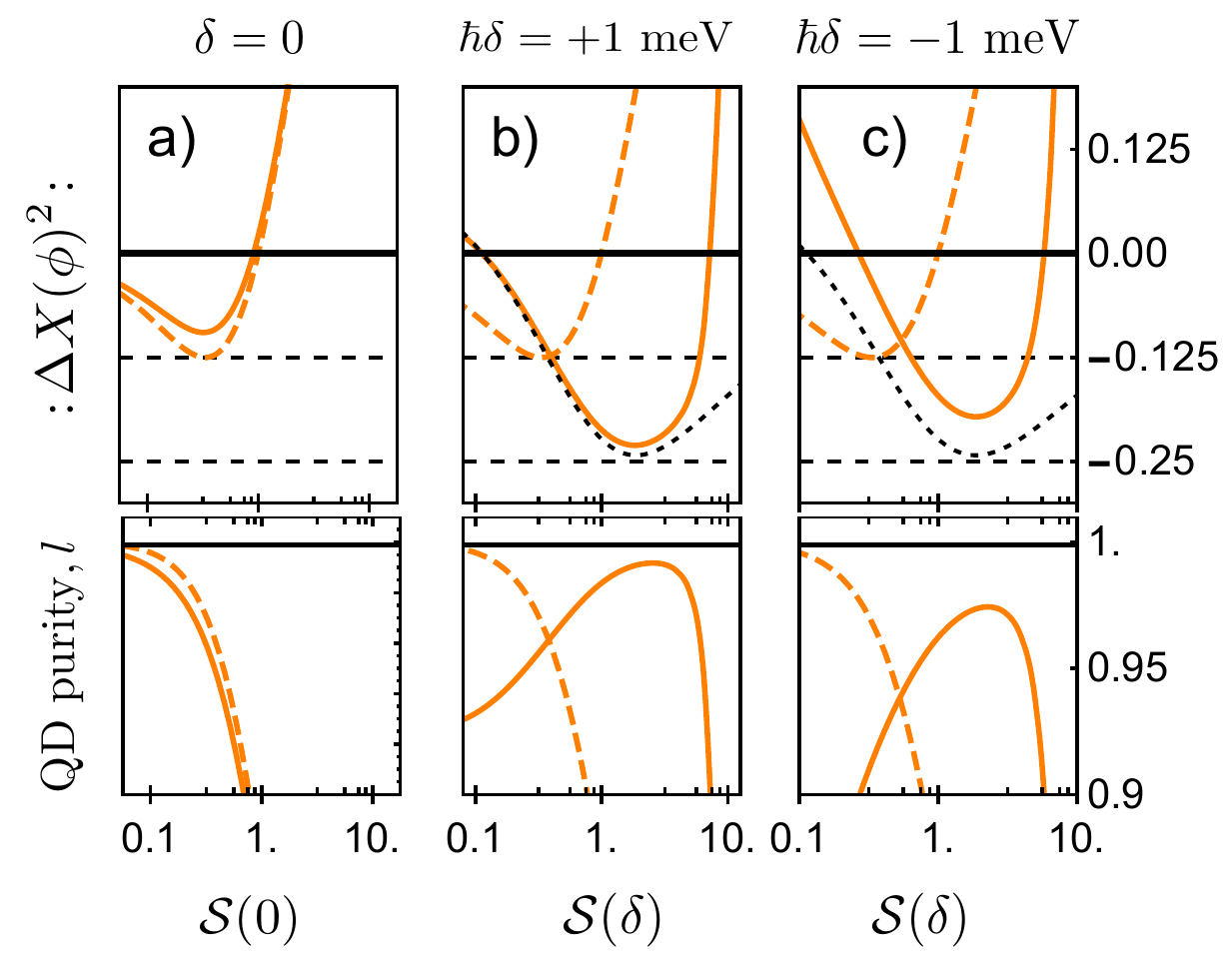}
	\caption{{\bf{Quadrature variance and exciton purity.}} 
	Normally ordered quadrature variance $\norm{\Delta X(\phi)^2}$ on (a) and off (b and c) resonance. 
	The solid curves are 
	calculated using the full phonon theory and the dashed curves are calculated in 
	the absence of phonons as in Eq.~({\ref{DeltaXWalls}}). The gridlines indicate the values $\norm{\Delta X(\phi)^2}=-0.125$ 
	(maximum level of squeezing obtainable in the absence of phonons) and $\norm{\Delta X(\phi)^2}=-0.25$ 
	(upper bound on the level of squeezing obtainable from a two-level system).  
	The black dotted curve shows the analytic approximation in Eq.~({\ref{DeltaXApprox}}). 
	The lower plots show the purity of the quantum dot excitonic state $l$, with 
	$l=1$ corresponding to a minimum uncertainty state of the emitted field. 
	Parameters are as in Fig.~\ref{qd_squeeze}.
Off-resonance ($\hbar\delta = \pm1~\mathrm{meV}$), the minimum and maximum scaled driving strengths cover the range $s\sim10^{5.5}-10^{7.5}$, while in the resonant case $\mathcal{S}(0)=s$ by definition.
 }
	\label{sections}
\end{figure}

Looking at the cases without phonons in Figs.~\ref{qd_squeeze} b) and d), 
we see that for all detunings the behaviour of the power contributions with 
increasing driving strength appears functionally the same, though with their maxima occurring 
at different driving strengths. 
This was seen in Eq.~({\ref{DeltaXWalls}}), which shows 
that the power contributions and quadrature variance depend only on the generalised saturation parameter $\mathcal{S}(\delta)$. 
When phonons are included, however, the situation appears more complex. 
On or near resonance for weak to moderate driving, the excitonic system is dominated by 
the spontaneous emission processes as it 
samples the phonon spectral density at the small generalised 
Rabi frequency $\sqrt{\Omega_\mathrm{r}^2+\delta_\mathrm{r}^2}$. 
The power contributions, and hence the quadrature variance, are then similar to that of an atomic system 
with no phonon coupling. This can be seen in Fig.~{\ref{sections}} a), where the quadrature 
variance is shown as a function of the saturation parameter 
on resonance, $\delta=0$, calculated with (solid curve) and without (dotted curve) phonons included. 
We see that $\norm{\Delta X(\phi)^2}$ has a minimum for $\mathcal{S}(0)=s=1/3$,  
in accordance with Eq.~({\ref{DeltaXWalls}}). This is the regime explored experimentally in Ref.~\cite{Schulte2015}. 
Although the phonons are not playing a qualitatively significant role here, 
we do see that the minimum of $\norm{\Delta X(\phi)^2}$ is slightly higher with them included. 
This can be attributed to the phonon sideband present in the QD emission spectrum, 
which acts to reduce the coherently scattered power below the level expected without phonons, 
even at low driving strengths~\cite{iles2016fundamental,mccutcheon2015optical,Iles-smith2017Nature}. 

\begin{figure*}
\begin{center}
	\includegraphics[width=0.95\textwidth]{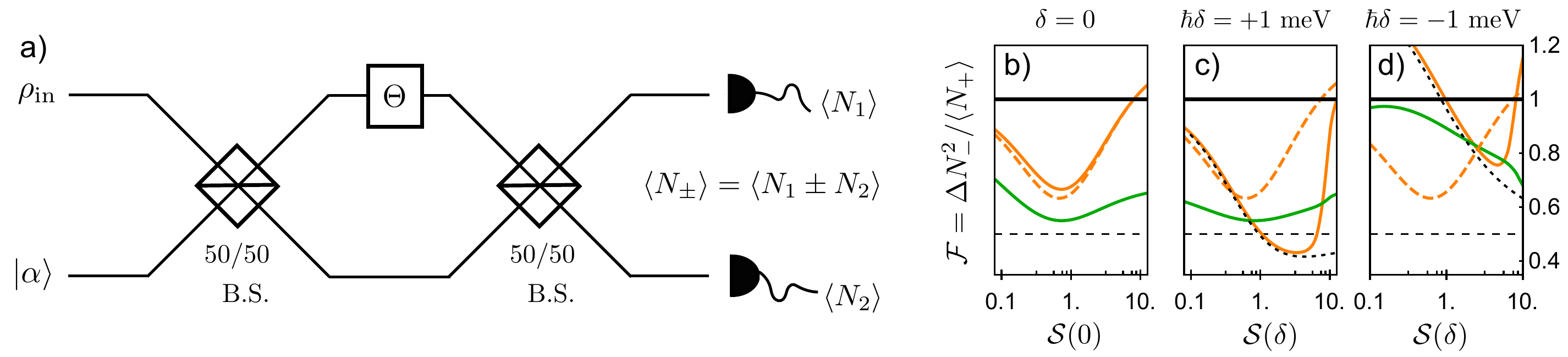}
	\caption{{\bf{Phase estimation with vibrationally enhanced resonance fluorescence fields.}} 
	a) Phase estimation setup, showing a test input state $\rho_{\mathrm{in}}$ in the first (upper)  
	mode and a coherent state input $\ket{\alpha}$ in the second (lower) mode of a Mach-Zehnder interferometer 
	consisting of two 50/50 beamsplitters separating and recombining the two 
	arms with path length difference parameterised by $\Theta$. When the first mode 
	is illuminated with resonance fluorescence light, the normalised phase sensitivity for $\Theta=\pi/2$ 
	is shown for resonant (b) and off-resonant driving (c and d), modelled with (solid, orange) and 
	without (dashed, orange) coupling to phonons. 
	Shown for reference is the normalised phase sensitivity for a squeezed vacuum input state in mode one, 
	with the squeezing parameter chosen such that the expectation value of the photon number 
	is equal to that in the resonance fluorescence case. The thick black line at $\mathcal{F}=1$ 
	is the phase sensitivity for a coherent state input in the first (upper) interferometer arm, 
	while the dashed black line at $\mathcal{F}=0.5$ is the global minimum for the 
	squeezed vacuum input. Parameters are as in Fig.~\ref{qd_squeeze} with $|\alpha|^2=1$. }
	\label{metrology}
\end{center}
\end{figure*}

Above saturation, we see that the power contributions with and without phonons 
markedly differ, as in this regime the phonon environment dominates over spontaneous emission due 
to its spectral density being sampled at larger generalised Rabi frequencies. The exciton then 
tends towards a thermal state as previously described, 
in which $P_{\mathrm{coh}}$ can be significant, while positive and negative detunings lead 
to a low and high total power $P$, respectively. 
As anticipated, this then gives rise to two regimes 
with quadrature squeezing, as shown explicitly in Figs.~{\ref{sections}} b) and c), 
where we plot the quadrature variance as a function of the generalised 
saturation parameter $\mathcal{S}$ for fixed detuning. 
The black dotted curves correspond to the approximate expression in 
Eq.~({\ref{DeltaXApprox}}), showing good agreement with the full phonon model until 
the driving strength becomes large enough 
that the decoupling regime is reached.
For positive detuning we see that a level of squeezing close to 
the upper bound $\norm{\Delta X(\phi)^2}=-0.25$ can be obtained, and that this is only possible  
when phonons are included. For the negative detuning case, as the driving increases phonons lead to thermalisation towards a state with total 
power $P>0.5$ due to steady-state population inversion 
[see Fig.~\ref{qd_squeeze} a)]~\cite{PhysRevLett.114.137401,PhysRevLett.107.193601,hughespopinv}. 
This means that the quadrature variance first increases at small driving, 
then begins to decrease with a discontinuous derivative (not shown) 
as $P$ passes through $0.5$. The  
strong driving exciton--phonon decoupling regime also sets in sooner than for positive detuning. 
Nevertheless, the obtained levels of squeezing still surpass 
those possible in the absence of the vibrational environment.

The lower plots in Fig.~\ref{sections} show the corresponding 
purity of the QD excitonic state parameterised by its Bloch vector length $1\geq l \geq 0$, 
which we use here to indicate whether the emitted field is a minimum uncertainty state (for which $l=1$). 
We see that when phonons are included, the regime of greatest squeezing coincides with 
the regime of greatest Bloch vector length, and that the optimal values of $l=1$ and $\norm{\Delta X(\phi)^2}=-0.25$ 
are closely approached in the positive detuning case.

We note that although the driving strengths required to reach the regime of vibrationally enhanced squeezing 
are experimentally challenging, they are not excessively so. Rabi energies of several hundred 
$\mu$eV are fairly routinely achieved experimentally~\cite{Unsleber:15}, and for the realistic parameters used in Fig.~{\ref{sections}}, 
the level of squeezing exceeds that possible in the absence of phonons for 
$\mathcal{S}(\delta)\approx 0.3$ corresponding to a Rabi energy of $\hbar\Omega\approx 0.8~\mathrm{meV}$, 
while the maximum squeezing occurs in the regime $\hbar\Omega\approx 1.5~\mathrm{meV}$. 
Broadly speaking, the vibrationally enhanced regime takes effect when $k_\mathrm{B} T < \hbar\sqrt{\Omega^2+\delta^2}$, 
and as such lowering the temperature below the $4~\mathrm{K}$ used here would allow 
it to be observed at lower driving strengths and detuning values. 
We emphasise that the vibrationally enhanced squeezing regime occurs when 
phonon-induced dephasing is large, and indeed dominates 
over spontaneous emission. As such, 
our results are robust against any additional pure-dephasing processes 
that are weak or moderate compared to the homogeneous linewidth of the exciton, 
as shown explicitly in Supplementary Note 4.
\\\\
\noindent
{\small{{\bf{Resonance fluorescence and phase estimation.}} }}
Having shown how a vibrational environment is able to lead to enhanced levels of quadrature squeezing in resonance 
fluorescence, a natural question to ask is whether the light produced is useful in applications. 
To address this question, we analyse the canonical phase estimation protocol illustrated in 
Fig.~{\ref{metrology}}. It consists of an unbalanced Mach-Zehnder interferometer with path difference 
parameterised by $\Theta$, and into which is inserted a pure coherent state $\ket{\alpha}$ in one arm 
and a general state $\rho_{\mathrm{in}}$ in the other arm. We are 
interested in the greatest accuracy with which $\Theta$ can be estimated. 
We consider perhaps the simplest measurement 
that can be used to construct an estimator for $\Theta$, 
which is the expectation value of the difference in photon numbers at the two outputs, 
$\langle N_- \rangle$, where $N_{\pm}=N_1\pm N_2$. 
Our accuracy figure of merit is then the variance in $N_-$, 
$ \Delta N_-^2 =\langle N_-^2 \rangle-\langle N_- \rangle^2$, normalised  
by the total photon flux $\langle N_+ \rangle$, which we label $\mathcal{F}= \Delta N_-^2/\langle N_+ \rangle$. 
Defined in this way, if a second coherent state is input into the first (upper) arm our 
figure of merit gives $\mathcal{F}=1$.

Considering first a squeezed single mode vacuum input state we 
recover the seminal result of Caves~\cite{PhysRevD.23.1693}. 
As detailed in the Methods, 
using $\rho_{\mathrm{in}} = \ketbra{\xi}{\xi}$ with 
$\ket{\xi}=S(\xi)\ket{\mathrm{vac}}$ and $S(\xi)=\exp[ (\xi a^2-\xi^* {a^{\dagger 2}})/2]$, 
we find $\langle N_+ \rangle = P_{\xi}+P_{\alpha}$ with 
$P_{\xi}=\sinh^2|\xi|$ and $P_{\alpha} = |\alpha|^2$, while 
the variance in the difference in photon numbers 
is minimised for $\Theta=\pi/2$, at which point 
\beq
\Delta N_-^2 = P_{\xi}+P_{\alpha}\e^{-2 |\xi|},
\label{SVvariance}
\eeq
showing that for $|\xi|>0$ the variance is reduced below that of a coherent 
state input with the same power~\cite{PhysRevD.23.1693,Lang2014}. 
For the squeezed vacuum we find $\mathcal{F}$ is minimised for 
$|\xi|=(1/2)\ln[1+2\sqrt{P_{\alpha}}]$ at which point $\mathcal{F}=1/(1+\sqrt{P_{\alpha}})$. 

For the resonance fluorescence field as the input state, 
for the same $\Theta$ the variance in $N_-$ is minimised 
when the phase of the coherent state amplitude $\alpha$ is equal to the dipole phase $\phi$, and is 
now given by 
\beq
 \Delta N_-^2 = P+P_{\alpha} (1- 4 P_{\mathrm{coh}}),
 \label{variance}
\eeq
with $P$ and $P_{\mathrm{coh}}$ the total and coherently 
scattered dimensionless powers as before.
Comparing 
Eqs.~({\ref{variance}}) and ({\ref{SVvariance}}), we can immediately 
see that if $P_{\mathrm{coh}}$ can reach values closest 
to its maximum of $0.25$, the second term in Eq.~({\ref{variance}}) will 
vanish, something that is only possible in the limit of infinite squeezing in the 
squeezed vacuum case.

Our figure of merit is shown in Fig.~{\ref{metrology}} as a function of driving strength, 
for a QD driven on b) and off-resonance [c) and d)] with (solid, orange curves) and 
without (dashed, orange) phonons, where we take $P_{\alpha}=1$. 
Shown also in green is the same figure of merit  
for the squeezed single mode vacuum, with the squeezing parameter $\xi$ chosen 
such that $P_{\xi} =P$, ensuring that the total photon flux is equal in both cases. 
The black dashed curve is calculated using the approximate 
QD steady-state in Eq.~({\ref{rhoth}}), showing good agreement with the full phonon model in the 
phonon enhanced regimes. 

Looking first at the resonant case b), we see that the figure of merit 
at low driving strengths is minimised by the squeezed vacuum input state, 
implying that for a fixed photon flux, it is this state which 
gives the greatest sensitivity to the phase $\Theta$. Hence, the resonance fluorescence 
field in the standard squeezing regime does not outperform the squeezed vacuum. 
Off-resonance and for positive detuning, however, we see that the resonance fluorescence field 
can out-perform the squeezed vacuum state, as in this regime 
$P_{\mathrm{coh}}\approx 0.25$ when phonons are included (c.f. Fig.~{\ref{qd_squeeze}}). 
Moreover, in this regime our figure of merit $\mathcal{F}$ reaches values below 
$0.5$, which is the minimum possible value for the squeezed vacuum 
for any parameters when $P_{\alpha}=1$. 
In the negative detuning case, although $P_{\mathrm{coh}}$ is again close $0.25$, the total 
power $P$ is now large, meaning the first term in Eq.~({\ref{variance}}) is significant. 
The competition between these two terms together with 
the phonon-induced population inversion gives rise to the complicated behaviour seen  
in Fig.~{\ref{metrology}} d). Interestingly, here 
we see that the resonance fluorescence field without phonons can actually 
outperform the squeezed vacuum, although the overall squeezed vacuum minimum of 
$\mathcal{F}=0.5$ is only beaten when including phonons in the positive detuning case. 
\\\\
\noindent
{\bf{\sffamily{Discussion}}}\\
\noindent
We have shown that interactions between a 
TLE and a vibrational environment can be harnessed 
to produce a source of quadrature squeezed light with levels of squeezing    
that would otherwise be impossible within resonance fluorescence. In fact, 
the obtainable squeezing can reach values very close to the fundamental bound 
for a two-level system. We have illustrated our findings with an 
explicit example of a QD coupled to phonons, which provides a 
feasible experimental platform 
to engineer such squeezed states of light.

We have also explored how these 
resonance fluorescence states perform in a phase estimation protocol. 
Without vibrational interactions, although the reduced fluctuations in 
resonance fluorescence states can provide an advantage over 
coherent state inputs, the single mode squeezed vacuum offers the 
overall best phase sensitivity when minimised over the squeezing magnitude. 
With vibrational interactions included, however, the resonance fluorescence state 
can outperform the squeezed vacuum, even when operating in its optimal regime, 
suggesting that resonance fluorescence fields could 
provide a useful resource in phase estimation applications. 
It is interesting to compare these findings with the Fisher information analysis of 
Ref.~\cite{Lang2014}, which states that for the setup 
shown in Fig.~{\ref{metrology}}, the optimal pure single mode input 
state over all phase estimators is the squeezed vacuum. 
Our findings therefore suggest that for more general multimode 
input states the squeezed vacuum ceases to be optimal, or that 
estimators beyond the difference in output photon numbers must be considered. 
As such, our results not only illustrate how vibrational environments 
can give rise to enhanced quadrature squeezing in resonance fluorescence, 
but also motivate future studies analysing the performance of generalised non-Gaussian multimode 
states in interferometry.
\\\\
\noindent
{\bf{\sffamily{Methods}}}\\
\noindent
{\small{{\bf{Quantum dot master equation.}}}} 
The QD couples to both vibrational and electromagnetic environments, which results in the total Hamiltonian 
$H = H_\mathrm{S} +H_{\mathrm{S-ph}} + H_{\mathrm{S-em}}+ H_\mathrm{B}$. 
At low temperatures the electron--phonon interaction is dominated by a linear displacement coupling with Hamiltonian 
$H_{\mathrm{S-ph}} = \sigma^\dagger\sigma\sum_k \hbar g_k(b_k^\dagger + b_k)$, where $b_k$ ($b_k^\dagger$) is the annihilation 
(creation) operator of phonon mode $k$ with frequency 
$\omega_k$ and coupling strength $g_k$~\cite{nazir2015modelling,Muljarov2004,Reigue2017,Gerhardt2018}. 
The electron--phonon coupling is characterised by the spectral density 
$J(\omega) = \sum_k\vert g_k\vert^2\delta(\omega-\omega_k)=
\alpha\omega^3\exp[-\omega^2/\omega_\mathrm{c}^2]$~\cite{nazir2015modelling}, with 
coupling strength $\alpha$ and 
cut-off frequency $\omega_\mathrm{c}$. 
Coupling to the electromagnetic field in the dipole and rotating wave approximations takes the form 
$H_{\mathrm{S-em}} = \hbar\sum_m h_m\sigma^\dagger a_m\e^{i \w_\mathrm{l} t} +\text{h.c.}$, 
where $a_m$ ($a_m^{\dagger}$) is the annihilation (creation) 
operator for mode $m$ of the field, with frequency $\nu_m$ and coupling strength $h_m$.
We assume the spectral density of the optical field varies slowly over the relevant energy scales of the system, allowing us to 
use the flat function $\sum_m\vert h_m\vert^2\delta(\nu - \nu_m)\approx 2\Gamma/\pi$, 
where $\Gamma$ is again the spontaneous emission rate. 
Free evolution of the environments is described by 
$H_\mathrm{B} = \hbar\sum_k \omega_k b_k^\dagger b_k +  \hbar\sum_m \nu_m a_m^\dagger a_m$. 

To account for the electron--phonon coupling we make use of the variational polaron 
transformation~\cite{nazir2015modelling,PhysRevB.84.081305,McCutcheon2013} defined by the unitary 
$\mathcal{U} = \ket{g}\bra{g} +  \ket{e}\bra{e}B_{+}$, where 
$B_\pm = \exp[\pm\sum_k f_k(b_k^\dagger - b_k)/\omega_k]$. 
This leads to a QD state dependent displacement of the phonon environment, where the $f_k$ are 
chosen to minimise the Feynman--Bogoliubov bound on the free energy, defining an optimised basis in which perturbation theory can then be applied. 
In the variational polaron frame we derive a second order Born-Markov master equation,  
which is valid for both strong and weak exciton--phonon coupling, as well as from weak to strong laser driving strengths. 
For real $\Omega$, this may be written compactly as
$
\dot\rho_\mathrm{V}= -(i/\hbar)\left[H_\mathrm{r},\rho_\mathrm{V}\right] 
+ \mathcal{K}_{\mathrm{ph}}[\rho_\mathrm{V}] + \Gamma\mathcal{L}_\sigma[\rho_\mathrm{V}],
$
where $\rho_\mathrm{V}$ is the reduced density operator of the 
QD in the variational frame and the renormalised system Hamiltonian is 
$H_\mathrm{r} = \hbar \delta_\mathrm{r}\sigma^{\dagger}\sigma+(\hbar\Omega_\mathrm{r}/2)(\sigma^{\dagger}+\sigma)$, 
with $\delta_\mathrm{r}$ and $\Omega_\mathrm{r}$ defined below. The phonon dissipator is 
\begin{align}
\mathcal{K}_{\mathrm{ph}}[\rho_\mathrm{V}]=&[\hat{\mathcal{Z}}_\mathrm{zz} \rho_\mathrm{V},\sigma^{\dagger}\sigma]\!-\!\frac{\Omega^2}{4}\big([\sigma_\mathrm{x},\hat{\mathcal{X}}_\mathrm{xx} \rho_\mathrm{V}]\!+\![\sigma_\mathrm{y},\hat{\mathcal{Y}}_\mathrm{yy} \rho_\mathrm{V}]\big)\nonumber\\
&+\frac{i \Omega}{2}\big([\sigma_\mathrm{y},\hat{\mathcal{Z}}_\mathrm{yz} \rho_\mathrm{V}]-[\sigma^{\dagger}\sigma,\hat{\mathcal{Y}}_\mathrm{yz} \rho_\mathrm{V}]\big)+\mathrm{h.c.}
\end{align}
where the rates are contained in the system operators 
\begin{align}
\hat{\mathcal{X}}_{\alpha\beta}&=\sum_{jk}\sigma_\mathrm{x}^{jk} \int\limits_0^{\infty}d\tau \e^{i\lambda_{jk}\tau}\Lambda_{\alpha\beta}(\tau)\ketbra{\psi_j}{\psi_k},\\
\hat{\mathcal{Y}}_{\alpha\beta}&=i\sum_{jk}\sigma_\mathrm{y}^{jk} \int\limits_0^{\infty}d\tau \e^{i\lambda_{jk}\tau}\Lambda_{\alpha\beta}(\tau)\ketbra{\psi_j}{\psi_k},\\
\hat{\mathcal{Z}}_{\alpha\beta}&=i\sum_{jk}(1+\sigma_\mathrm{z}^{jk}) \int\limits_0^{\infty}d\tau \e^{i\lambda_{jk}\tau}\Lambda_{\alpha\beta}(\tau)\ketbra{\psi_j}{\psi_k},
\end{align}
which are written in terms of the phonon correlation functions 
$\Lambda_{xx}(\tau)=(B^2/2)(\e^{\kappa(\tau)}+\e^{-\kappa(\tau)}-2)$, 
$\Lambda_{yy}(\tau)=(B^2/2)(\e^{\kappa(\tau)}-\e^{-\kappa(\tau)})$ and 
$\Lambda_{zz}(\tau)=\int_0^{\infty} d \w J(\w) (1-F(\w))^2C_{\parallel}(\tau,\omega)$, with 
$\kappa(\tau)=\int_0^{\infty}d\w J(\w)F(\w)^2\w^{-2}C_{\parallel}(\tau,\omega)$ and 
\beq
C_{\parallel}(\tau,\omega)=\coth\Big(\frac{\hbar\beta \w}{2}\Big)\cos(\w\tau)-i \sin(\w\tau),
\eeq
while $\Lambda_{yz}(\tau)=-2 B \int_0^{\infty} d \w J(\w) \w^{-1} F(\w)(1-F(\w))C_{\perp}(\tau,\omega)$ with 
\beq
C_{\perp}(\tau,\omega)=\coth\Big(\frac{\hbar\beta \w}{2}\Big)\sin(\w\tau)+i \cos(\w\tau).
\eeq
The eigenstates of the renormalised system Hamiltonian satisfy 
$H_\mathrm{r}\ket{\psi_j}=\psi_j\ket{\psi_j}$, 
and we have defined $\hbar\lambda_{jk}=\psi_j-\psi_k$ and $\sigma_\alpha^{jk}=\bra{\psi_j}\sigma_{\alpha}\ket{\psi_k}$. 
In the above the displacement operator thermal average is 
$
B=\langle B_{\pm}\rangle=\exp[-\kappa(0)/2]
$
and $F(\w)$ is the variationally determined mode displacement defined in Supplementary Note 1.
The renormalised detuning and Rabi frequency are 
$\delta_\mathrm{r} = \delta+\int_0^{\infty} d\w J(\w)\w^{-1}F(\w)(F(\w)-2)$ and 
$\Omega_\mathrm{r}=\Omega B$. 

We note that in the variational frame the dipole operator carries a displacement operator, such that $\sigma \to B_-\sigma$.
This displacement operator leads to a phonon sideband, a 
consequence of non-Markovian lattice relaxation during the emission 
process~\cite{iles2016fundamental,mccutcheon2015optical,Iles-smith2017Nature,PhysRevB.92.205406}. 
Including this effect, the field emitted by the QD becomes 
${E}^{(+)}(t) = {E}_0(t) -\sqrt{2\Gamma/\pi}B_-(t)\sigma(t)$, and 
we obtain a modification to the quadrature variance,  
which becomes [cf.~Eq.~({\ref{NormXVar})] 
$
\norm{\Delta X(\phi)^2}=1-|2P-1|-4B^2P_{\mathrm{coh}}.
$
Similarly, Eq.~({\ref{variance}}) becomes $\Delta N_-^2 = P+P_{\alpha} (1- 4 B^2P_{\mathrm{coh}})$.
\\\\
\noindent
{\small{{\bf{Phase estimation protocol.}}}}
The phase estimation setup we consider is shown and described in Fig.~\ref{metrology}, 
and relies on an estimator based on the difference in intensities at the two outputs, 
which has corresponding dimensionless operator $N_-=N_1-N_2$. 
As such we seek to analyse the variance in this quantity, 
$\Delta N_-=\langle N_-^2\rangle - \langle N_- \rangle^2$. 
To proceed we write the output number operators in the steady state as 
$N_1 = \smash{E_1^{(-)}E_1^{(+)}}$ with 
$E_1^{(+)}$ the positive frequency component of the electric field operator in arm $1$, 
and similarly for arm $2$. Working in the Heisenberg picture and neglecting 
retardation effects, we can relate these operators at the outputs to those at the inputs using 
\beq
\left(
\begin{array}{c}
E^{(+)}_1 \\
E^{(+)}_2 \end{array}\right)
= \frac{1}{2}\left(\begin{array}{c}
E^{(+)}_-(0)+E^{(+)}_+(\Theta) \\
iE^{(+)}_-(0)-iE^{(+)}_+(\Theta) \end{array}\right),
\eeq
where $E^{(+)}_{\pm}(\Theta)=E_{\mathrm{in}}^{(+)}(\Theta)\pm i E_{\alpha}^{(+)}(\Theta)$, 
with $E_{\mathrm{in}}^{(+)}(\Theta)$ the positive frequency component 
of the electric field at the first input arm illuminated by the state $\rho_{\mathrm{in}}$, 
propagated by the path difference 
parameter angle $\Theta$, while $E_{\alpha}^{(+)}(\Theta)$ is the corresponding 
quantity at the second input arm illuminated by the coherent state $\ket{\alpha}$. 
Using these relations we straightforwardly find 
$N_-=(1/2)[E_-^{(-)}(0)E_+^{(+)}(\Theta)+E_+^{(-)}(\Theta)E_-^{(+)}(0)]$, 
and where at this stage we have made no assumptions about the state 
of the field or the path length difference $\Theta$. 

The second arm is illuminated by a single mode coherent state. As such, when expectation values 
are taken, all other modes do not contribute, and neglecting constants we can therefore write 
the dimensionless electric field operator as 
$E_{\alpha}^{(+)}(\Theta)= E_{\alpha}^{(+)}(0)\e^{-i \Theta}=a_2\e^{-i \Theta}$. 
A similar argument holds when the first arm is illuminated by the single mode squeezed vacuum. 
In the resonance fluorescence case the situation is more subtle. Omitting the free field (vacuum) contribution, 
and again dropping constants, we have 
$\smash{E^{(+)}_{\mathrm{in}}(\Theta)=\sigma_\mathrm{L}(\tau_{\Theta})}$, where 
the subscript signifies a `lab'-frame (non-rotating) operator and 
$\tau_{\Theta}$ is the time delay corresponding to the phase shift $\Theta$. 
In the multimode resonance fluorescence case it is not obvious that a time delay gives 
rise to simple phase factor on the electric field operators. However, 
since we work in a rotating frame, we can write 
$\sigma_\mathrm{L}(\tau_{\Theta})=\sigma(\tau_\Theta)\e^{-i \tau_\Theta \w_\mathrm{l}}$ 
where $\sigma$ is the rotating frame dipole operator and $\w_\mathrm{l}$ is the driving laser frequency. 
If we are interested in delays $\tau_\Theta$ that give rise to phase shifts $\Theta=\tau_\Theta \w_\mathrm{l}\approx \pi$, 
this corresponds to $\tau_\Theta \sim \mathrm{fs}$ for $\w_\mathrm{l}\sim 1~\mathrm{eV}$. Over these 
very short times the rotating frame operator $\sigma$ does not significantly change, 
and we can write $E^{(+)}_{\mathrm{in}}(\Theta)=E^{(+)}_{\mathrm{in}}(0)\e^{-i\Theta}=\sigma(0)\e^{-i\Theta}$. 

Putting these results together, in the case of the squeezed vacuum input, for which 
we have $\rho_{\mathrm{in}}=\ketbra{\xi}{\xi}$ with 
$\ket{\xi}=\exp[(\xi a^2-\xi^* a^{\dagger 2})/2]\ket{\mathrm{vac}}$, 
we find 
\beq
\langle N_- \rangle = \cos\Theta (P_\xi - P_\alpha),
\eeq
with $P_{\xi}=\sinh^2|\xi|$ the (dimensionless) power in the squeezed vacuum state, 
and $P_{\alpha}=|\alpha|^2$ the power in the coherent state. The variance in the 
difference in photon number for this input state is 
\begin{align}
&\Delta N_-^2 = P_{\alpha}+P_\xi\big(P_\xi+3/2\big)+\cos(2\Theta)P_\xi\big(P_\xi+1/2\big)\nonumber\\
&+(1-\cos(2\Theta))P_\alpha\Big(P_\xi+\sqrt{P_{\xi}(P_\xi+1)}\cos\Delta\phi\Big),
\end{align}
where we have written $\alpha=|\alpha|\e^{i\phi_{\alpha}}$ and 
$\xi=|\xi|\e^{-i\phi_\xi}$ and $\Delta\phi=2\phi_{\alpha}-\phi_{\xi}$. 
This expression is minimised for $\Theta=\pi/2$ and $\Delta\phi=\pi$, at which point we have 
\begin{align}
\Delta N_-^2 =& P_\xi+P_{\alpha}\Big(1+2P_\xi-2\sqrt{P_\xi(P_\xi+1)}\Big),\nonumber\\
=&P_\xi+P_\alpha \e^{-2 |\xi|},
\end{align}
which is known previously from the work of Caves~\cite{PhysRevD.23.1693}. 

In the resonance fluorescence case we use $\Theta=\pi/2$ to facilitate a 
far comparison, and now find 
\beq
\Delta N_-^2 = P+P_\alpha (1-4 P_{\mathrm{coh}} \cos^2\Delta\phi),
\eeq
where now $P=\langle \sigma^{\dagger}\sigma\rangle$ is the (dimensionless) power in 
the resonance fluorescence field, and we have written 
$\langle \sigma \rangle = |\langle \sigma \rangle| \e^{-i\phi}=\sqrt{P_{\mathrm{coh}}}\e^{-i \phi}$ 
with $\phi$ the dipole phase in the steady-state and $\Delta\phi=\phi_{\alpha}-\phi$. In this 
case the minimum corresponds to $\Delta \phi=0$, and with this substitution we arrive 
at Eq.~({\ref{variance}}).  
\\\\
\noindent
{\bf{\sffamily{Data availability}}}\\
\noindent
The data that support the findings of this study are available from the corresponding authors upon reasonable request.

\providecommand{\noopsort}[1]{}\providecommand{\singleletter}[1]{#1}%

\vspace{5mm}
{\bf{\sffamily{Acknowledgements}}}\\
\noindent
The authors wish to thank Alistair Brash, Pieter Kok and Jonathan Matthews for useful discussions. 
This project has received funding from the 
European Union's Horizon 2020 research and innovation programme under the 
Marie Sk{\l}odowska-Curie grant agreement No. 703193. A.N.~is supported by the EPSRC, grant no.~EP/N008154/1, 
and J.I.-S. acknowledges support from the Royal Commission for the Exhibition of 1851.\\

\noindent
{\bf{\sffamily{Author Contributions}}}\\
\noindent
JIS derived the underlying formalism and master equation. 
DPSM derived the main results, with input from all authors. 
JIS, AN, and DPSM all contributed towards analysis, discussions and preparation of the manuscript.\\

\noindent
{\bf{\sffamily{Competing Financial Interests}}}\\
\noindent
The authors declare no competing interests.\\
\newpage

\clearpage

\widetext
\section*{\large Supplemental Material}

\section{Supplementary Note 1 - Variational polaron transformation}

The starting point for our quantum dot model is a Hamiltonian describing a two level system (TLS) with 
ground and excited states $\ket{g}$ and $\ket{e}$ respectively, driven by a classical 
laser field with frequency $\omega_\mathrm{l}$ and Rabi frequency $\Omega$. 
As introduced in the main text we have, 
\begin{equation}
H = \hbar\delta\sigma^\dagger\sigma+ \frac{\hbar\Omega}{2}\sigma_\mathrm{x} + \hbar\sigma^\dagger\sigma\sum\limits_k g_k(b^\dagger_k + b_k) + \hbar\sum\limits_m \left(h_m \sigma^\dagger a_m e^{i\omega_\mathrm{l} t} + \text{h.c.}\right)
+\sum\limits_k \hbar\w_k b^\dagger_k b_k + \sum\limits_m\hbar \nu_m a^\dagger_m a_m,
\end{equation}
where $\sigma=\ketbra{g}{e}$ and $\sigma_\mathrm{x}=\sigma+\sigma^{\dagger}$. 
The Hamiltonian is written in a frame rotating with respect to the laser frequency $\omega_\mathrm{l}$ 
and in the rotating wave approximation, with the laser and QD transition detuned by $\delta = \omega_0 - \omega_\mathrm{l}$. 
To model the dynamics of the TLS, we apply the variational polaron transformation to the above Hamiltonian, 
allowing us to derive a master equation valid outside of the weak electron-phonon coupling regime~\cite{McCutcheon2013,PhysRevB.84.081305,nazir2015modelling}.
The variational polaron transformation is given by the state dependent displacement operator
\begin{equation}
\mathcal{U}_\mathrm{V} = \ket{g}\!\bra{g}  + \ket{e}\!\bra{e}B_+,
\end{equation}
where $B_\pm = \exp\big[\pm\sum_kf_k(b_k^\dagger - b_k)/\w_k \big]$ is a multi-mode displacement operator, and $f_k$ is a parameter which we shall use to variationally optimise the transformation in the next section.

Applying this transformation to the initial Hamiltonian yields 
$H_\mathrm{V}= \mathcal{U}_\mathrm{V}  H\mathcal{U}^\dagger_\mathrm{V} = H_\mathrm{r}+ H_\mathrm{I}^{\mathrm{em}} + H_\mathrm{I}^{\mathrm{ph}}+ H_\mathrm{B}$, where 
$H_\mathrm{r} = \hbar\delta_\mathrm{r}\sigma^{\dagger}\sigma + (\hbar\Omega_\mathrm{r}/2)\sigma_\mathrm{x}$, 
and $H_\mathrm{B} = H_\mathrm{B}^{\mathrm{em}} + H_\mathrm{B}^{\mathrm{ph}}$, with 
$H_\mathrm{B}^{\mathrm{ph}}=\hbar\sum_k\w_k b_k^\dagger b_k$ and $ H_\mathrm{B}^{\mathrm{em}}=\hbar\sum_m\nu_m a^\dagger_m a_m$. 
Notice that we have introduced the renormalised detuning 
$\delta_\mathrm{r} = \delta + R$, with $R = \sum_k\omega_k^{-1}f_k(f_k-2g_k)$, 
and renormalised driving strength $\Omega_\mathrm{r}=\Omega B$, where 
$ B=\langle B_{\pm} \rangle_{H_\mathrm{B}}= \tr_\mathrm{B}(B_\pm\rho_\mathrm{B}^{\mathrm{ph}})$ is the expectation value of the displacement operator with respect to the thermal state 
$\rho_\mathrm{B}^{\mathrm{ph}} = \exp\big(-\beta H_\mathrm{B}^{\mathrm{ph}} \big)
/\tr_\mathrm{B}\big[\exp(-\beta H_\mathrm{B}^{\mathrm{ph}})\big]$ with $\beta = 1/k_\mathrm{B} T$ the inverse temperature. 
The transformed interaction Hamiltonians read 
\begin{align}
&	H_\mathrm{I}^{\mathrm{ph}} = 
\frac{\hbar\Omega}{2}\left(\sigma_\mathrm{x} B_\mathrm{x} + \sigma_\mathrm{y} B_\mathrm{y}\right)  +\hbar \sigma^\dagger\sigma B_\mathrm{z}
\hspace{0.5cm}\text{and}\hspace{0.5cm}
H_\mathrm{I}^{\mathrm{em}} =\hbar \sum_m h_m \sigma^\dagger B_+ a_m e^{i\omega_\mathrm{l}t} +\mathrm{h.c.}
\end{align}
where $B_\mathrm{x} = (B_+ + B_-  -2 B)/2$, $B_\mathrm{y} = i(B_+ - B_-)/2$, and $B_\mathrm{z} = \sum_k(g_k - f_k)(b_k^\dagger + b_k)$. 
Notice that the electron--phonon interaction now contains a mixture of displacement operators and linear coupling. 
In the next section we outline how the variational principle can be used to specify the contribution of each of these terms.

\subsection*{Minimising the free energy}

As mentioned above, we now choose the displacement such that the transformed Hamiltonian minimises the free energy of the system. 
We do so by minimising the Feynman--Bogliubov upper bound on the free energy~\cite{nazir2015modelling}, that is:
$$
A_\mathrm{B} = -\frac{1}{\beta}\ln\left[\tr(e^{-\beta H_0})\right] + \langle H_\mathrm{I}^{\mathrm{ph}}\rangle_{H_0} + \mathcal{O}\left( \langle H_\mathrm{I}^2\rangle_{H_0}\right) ,
$$
where $H_0 = H_\mathrm{r} + H_\mathrm{B}$. 
This procedure allows us to derive a master equation which is valid over a broad range of parameters, and does not suffer from the same pathologies as 
standard polaron theory (see Refs.~\cite{PhysRevB.84.081305} and \cite{nazir2015modelling} for a detailed discussion).
Thus, minimising $A_\mathrm{B}$ with respect to $f_k$, we obtain the expression
\begin{equation}
\frac{\partial A_\mathrm{B}}{\partial f_k}= \frac{1}{\tr\left[e^{-\beta H_0}\right]}\tr\left[\frac{\partial H_0}{\partial f_k} e^{-\beta H_0}\right]=0.
\end{equation} 
Solving this equation, and substituting the minimised displacements into the expressions for the renormalised system parameters, 
we find that in the continuum limit we have:
\begin{align}
\Omega_\mathrm{r} &= \Omega B=\Omega\exp\left[-\frac{1}{2}\int_0^\infty \frac{J(\w)F^2(\w)}{\w^2}\coth\left(\frac{\hbar\beta\w}{2}\right)d\w\right]\hspace{0.5cm}\text{and}\hspace{0.5cm}
\delta_\mathrm{r} =\delta +\int_0^\infty \frac{J(\w)F(\w)}{\w}(2-F(\w))d\w,
\end{align}
where we have introduced the phonon spectral density $J(\w)=\sum_k |g_k|^2 \delta (\w-\w_k)$ and the variational function:
\begin{equation}
F(\w) = \frac{\eta_\mathrm{r} - \delta_\mathrm{r}\tanh\left(\hbar\beta\eta_\mathrm{r}/2\right)}{\eta_\mathrm{r}-\tanh\left(\hbar\beta\eta_\mathrm{r}/2\right)\left(\delta_\mathrm{r} - \frac{\Omega_\mathrm{r}^2}{2\w}\coth\left(\hbar\beta\w/2\right)\right)}.
\end{equation}
Here $\eta_\mathrm{r} = \sqrt{\delta_\mathrm{r}^2 + \Omega_\mathrm{r}^2}$ is the renormalised generalised Rabi frequency.
These equations can be solved self-consistently to find the renormalised parameters that minimise the Feynman--Bogliubov free energy.

\section*{Supplementary Note 2 - Variational master equation}

To describe the dynamics of the reduced state of the TLS, $\rho_\mathrm{V}(t)$, we shall treat the interaction Hamiltonian $H_\mathrm{I} = H_\mathrm{I}^{{\rm {\rm \mathrm{em}}}} + H_\mathrm{I}^{{\rm {\rm \mathrm{ph}}}}$ 
to second order using a Born-Markov master equation in the variational polaron frame, 
which in the interaction picture takes the form~\cite{breuer2007theory}:
\begin{equation} 
\frac{\partial\tilde\rho_\mathrm{V}(t)}{\partial t}=-\frac{1}{\hbar^2}\int\limits_0^\infty d\tau ~\tr_\mathrm{B}\left[H_\mathrm{I}(t),\left[H_\mathrm{I}(t-\tau), \tilde\rho_\mathrm{V}(t)\otimes\rho_\mathrm{B}^{{\rm \mathrm{em}}}\otimes\rho_\mathrm{B}^{{\rm {\rm \mathrm{ph}}}}\right]\right],
\end{equation}
where $\tilde{\rho}_\mathrm{V}(t)=\exp[i H_0 t]\rho_\mathrm{V}(t)\exp[-i H_0 t]$ and $H_\mathrm{I}(t)=\exp[i H_0 t]H_\mathrm{I}\exp[-i H_0 t]$ are interaction picture operators, 
and we have made the Born approximation which is to factorise the environmental density operators, here in the variational polaron frame, 
such that they remain 
static throughout the evolution of the system. Note that correlations may be generated between the system and the phonon environment in the original representation. 
We shall assume that in the variational polaron frame the phonon environment remains in the thermal state defined above, while the electromagnetic environment remains in its vacuum state $\rho_\mathrm{B}^{{\rm \mathrm{em}}} = \bigotimes_m \ket{0_m}\bra{0_m}$.
Since the trace over the chosen states of the environments removes terms linear in creation and annihilation operators, we may split the master equation into two separate contributions corresponding to the phonon and photon baths respectively~\cite{mccutcheon2015optical},
\begin{equation}
\frac{\partial\tilde\rho_\mathrm{V}(t)}{\partial t}= \mathcal{K}_{{\rm {\rm{ph}}}}[\tilde\rho_\mathrm{V}(t)] + \mathcal{K}_{{\rm{em}}}[\tilde\rho_\mathrm{V}(t)].
\end{equation}
In the subsequent sections we shall analyse each of these contributions in turn.

\subsection*{Phonon contribution}

To derive the contribution from the phonon environment, we follow Ref.~\cite{McCutcheon2013}.
We start by transforming into the interaction picture with respect to the Hamiltonian 
$H_0=\frac{\hbar\delta_\mathrm{r}}{2} \sigma_\mathrm{z} + \frac{\hbar\Omega_\mathrm{r}}{2}\sigma_\mathrm{x}  
+ \frac{\hbar R}{2}\id+ \hbar\sum_k\omega_k b_k^\dagger b_k + \hbar\sum_m \nu_m a_m^\dagger a_m$. 
Using this transformation, the phonon component of the interaction Hamiltonian takes the form
\begin{equation}
H^{{\rm {\rm \mathrm{ph}}}}_\mathrm{I}(t)= \frac{\hbar\Omega}{2}\left(\sigma_\mathrm{x}(t) B_\mathrm{x}(t) + 
\sigma_\mathrm{y}(t) B_\mathrm{y}(t)\right) + \hbar\sigma^\dagger\sigma(t) B_\mathrm{z}(t).
\end{equation}
Here $B_\mathrm{x} (t) =\frac{1}{2} (B_+(t) + B_-(t) - 2B)$, $B_\mathrm{y}(t) = \frac{i}{2}(B_+(t)-B_-(t))$, and 
$B_\mathrm{z}(t) = \sum_k(g_k-f_k)(b_k e^{-i\omega_k t} + b_k^\dagger e^{i\omega_k t})$, where 
$B_\pm(t) = \exp[\pm\sum_k\frac{f_k}{\omega_k} (b^\dagger_k e^{i\nu_kt} -b_ke^{-i\nu_kt})]$. 
The system operators can be formally written in the interaction picture as 
$
\sigma_\alpha(t) = \sum_{jk} \sigma_{\alpha}^{jk} e^{i\lambda_{jk} t}\ket{\psi_j}\!\bra{\psi_k},
$
where $\ket{\psi_j}$ are the eigenstates of the renormalised system Hamiltonian satisfying $H_\mathrm{r}\ket{\psi_j} =\psi_j\ket{\psi_j}$, 
$\hbar\lambda_{jk}=\psi_j - \psi_k$, and $\sigma_\alpha^{ij} = \bra{\psi_i}\sigma_\alpha\ket{\psi_j}$ with $\alpha \in\{\mathrm{x,y,z}\}$. 
Using these expressions and moving back into the Schr\"{o}dinger picture we arrive at 
the form of $K_{\mathrm{ph}}$ given in the main text.

\subsection*{Photon contribution}

We now focus on the interaction between the electromagnetic field and the TLS. 
The interaction picture Hamiltonian for the field may be written as 
$H_\mathrm{I}^{{\rm \mathrm{em}}}(t)= \sigma^\dagger(t) e^{i\omega_\mathrm{l} t} B_+(t) A(t)  + \text{h.c.}$, where 
$A(t) = \hbar\sum_m h_m {a_m} e^{-i\omega_m t}$ and $B_+(t)$ is as given in the previous section. 
If we consider the interaction picture transformation for the system operators we have
\begin{equation}
\sigma(t)\mathrm{e}^{-i\omega_\mathrm{l}t} = \exp\left[i\left(\frac{\delta_\mathrm{r}}{2}\sigma_\mathrm{z} + \frac{\Omega_\mathrm{r}}{2}\sigma_\mathrm{x}\right) t \right]
\sigma \exp\left[-i\left(\frac{\delta_\mathrm{r}}{2}\sigma_\mathrm{z} + \frac{\Omega_\mathrm{r}}{2}\sigma_\mathrm{x}\right) t \right]\mathrm{e}^{-i\omega_\mathrm{l}t}
\approx \sigma e^{-i\omega_0 t},
\end{equation}
where we have used the fact that $\omega_\mathrm{l}\gg\Omega_\mathrm{r},~\delta_\mathrm{r}$ for typical solid-state emitters to simplify the interaction picture transformation~\cite{carmichael1998statistical,McCutcheon2013}. 
By substituting this expression into the photon contribution of the master equation and recalling that all modes of the field are in their vacuum state, 
we have
\begin{equation}
\mathcal{K}_{{\rm \mathrm{em}}}[\tilde\rho_\mathrm{V}(t)] = -\frac{1}{\hbar^2}\int\limits_0^\infty d\tau ~\text{tr}_\mathrm{B}\left[H^{{\rm \mathrm{em}}}_\mathrm{I}(t),\left[H^{{\rm \mathrm{em}}}_\mathrm{I}(t-\tau), \tilde\rho_\mathrm{V}(t)\otimes\rho_\mathrm{B}^{{\rm \mathrm{em}}}\right]\right] = \gamma(\omega_0)\left(\sigma\tilde\rho_\mathrm{V}(t)\sigma^\dagger-(1/2)\left\{\sigma^\dagger\sigma,\tilde\rho_\mathrm{V}(t)\right\}\right),
\end{equation}
with the anticommutator $\{A,B\}=AB+BA$, and where the spontaneous emission rate is given 
by~\cite{McCutcheon2013,roy2015spontaneous,PhysRevB.92.205406}
\begin{equation}
\gamma(\omega_0) = \text{Re}\left[\int_0^\infty e^{i\omega_0\tau}G(\tau) \Lambda(\tau)d\tau\right].
\end{equation}
Here $\Lambda(\tau) = \int_0^\infty d\nu J_{{\rm \mathrm{em}}}(\nu)e^{i\nu\tau}$, with 
$J_{{\rm \mathrm{em}}}(\nu) = \sum_m\vert f_m\vert^2\delta(\nu - \nu_m)$ 
being the spectral density of the electromagnetic environment, 
and $G(\tau)$ is a phonon correlation function given by 
\beq
G(\tau)=B^2 \mathrm{exp}\Big[\int_0^\infty \frac{J(\w)F(\w)^2}{\w^2}\Big(\operatorname{coth}(\hbar\beta\w/2)\cos{\w \tau}+i\sin{\w t}\Big)d\nu\Big].
\eeq
As discussed in the manuscript, the local density of states of the electromagnetic 
field does not vary appreciably over energy scales relevant to QD systems in bulk, 
which allows us to make the assumption that the spectral density is 
flat~\cite{carmichael1998statistical,McCutcheon2013}, $J_{{\rm \mathrm{em}}}(\nu)\approx2\Gamma/\pi$. 
The electromagnetic correlation function may then be evaluated as $\Lambda(\tau)\approx  \Gamma\delta(\nu)+i\mathcal{P}[1/\tau]$, 
where $\mathcal{P}$ denotes the principal value integral.
Combining these expressions and resolving the remaining integral, we find that the spontaneous emission rate takes on the form 
$\gamma(\omega_0) \approx \Gamma$, where we have used the fact that $G(0) = 1$, such that in the Schr\"{o}dinger picture we have
\begin{equation}
\mathcal{K}_{{\rm \mathrm{em}}}[\rho_\mathrm{V}(t)] = \Gamma\mathcal{L}_\sigma\left[\rho_\mathrm{V}(t)\right] = \Gamma\left(\sigma\rho_\mathrm{V}(t)\sigma^\dagger-(1/2)\left\{\sigma^\dagger\sigma,\rho_\mathrm{V}(t)\right\}\right).
\end{equation}

Although for a flat optical spectral density the variational polaron transformation does not alter the spontaneous emission rate, it does directly influence the first-order correlation function, and therefore the emission spectrum of the system.
As discussed in detail in Refs.~\cite{iles2016fundamental,Iles-smith2017Nature}, 
when calculated in the variational polaron frame, the steady-state first-order correlation function takes the form,
\begin{equation}
g^{(1)}(\tau) = G(\tau)\langle\sigma^\dagger(\tau)\sigma\rangle_\mathrm{V},
\label{g1var}
\end{equation}
where the first factor is the phonon correlation function, which leads to the emergence of a 
non-Markovian phonon sideband in the emission spectrum~\cite{iles2016fundamental,Iles-smith2017Nature,roy2015spontaneous,PhysRevB.92.205406}.
The second term is the first-order correlation function calculated in the variational polaron frame, and describes pure optical emission processes from the TLS.
The phonon sideband contribution is particularly important for the coherent scattered power, 
which is given by the long-time limit of Eq.~({\ref{g1var}}). That is, 
we have $P_{\mathrm{coh}}=\lim_{\tau\to\infty} g^{(1)}(\tau) = |\langle \sigma \rangle |^2=B^2 |\langle \sigma \rangle_\mathrm{V}|^2$, 
with the prefactor $B^2$ acting to decrease the overall fraction of coherently scattered light~\cite{iles2016fundamental}.

\begin{figure*}
	\begin{center}
		\includegraphics[width=0.45\textwidth]{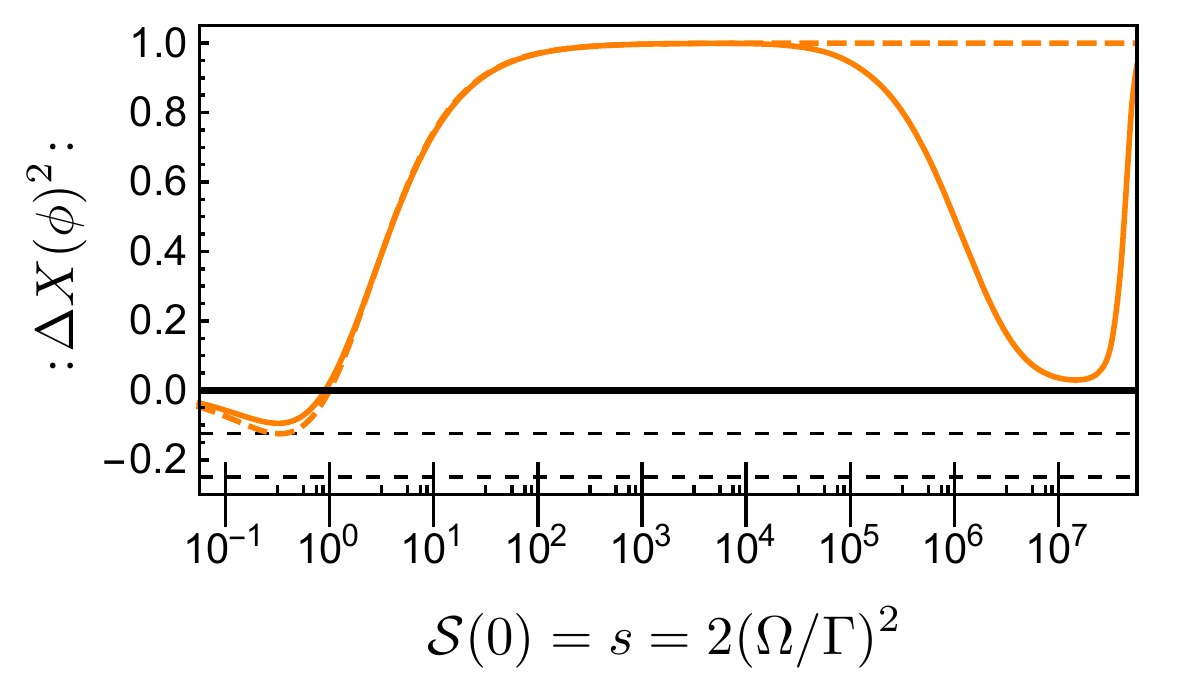}
		\caption{Quadrature variance for resonant excitation $\delta=0$, plotted as a function of driving strength from well below 
			to well above saturation. The phonon enhanced coherent scattering regime can be seen as a decrease in the 
			quadrature variance around $\mathrm{S}(0)\approx 10^5$, though this does not giving rise to quadrature squeezing as 
			the normally ordered variance remains positive.}
		\label{FullResonant}
	\end{center}
\end{figure*}

\section*{Supplementary Note 3 - Quadrature variance in the resonant phonon-enhanced scattering regime}

As explained in the main text, quadrature squeezing in resonance fluorescence occurs when the total 
power $P$ and the coherently scattered power $P_{\mathrm{coh}}$ satisfy the condition 
$:\Delta X(\phi)^2:=1-|2P-1|-4P_{\mathrm{coh}}<0$. As such, it makes sense to first explore 
the regime of phonon-enhanced coherently scattering 
identified in Ref.~\cite{McCutcheon2013}, in which $P_\mathrm{coh}$ was shown to take on significant values 
in the strong driving regime on resonance. As can be seen in Supplementary Figure~\ref{FullResonant}, however, 
on resonance, no squeezing occurs above saturation, although a reduction in the quadrature variance is apparent. 
This is because in this regime the coherent and incoherent contributions become approximately equal, 
and we have $P=0.5$ and $P_\mathrm{coh}=0.25$, resulting in $:\Delta X(\phi)^2:\approx 0$. 

\section*{Supplementary Note 4 - Additional dephasing effects}

\begin{figure*}
	\begin{center}
		\includegraphics[width=0.45\textwidth]{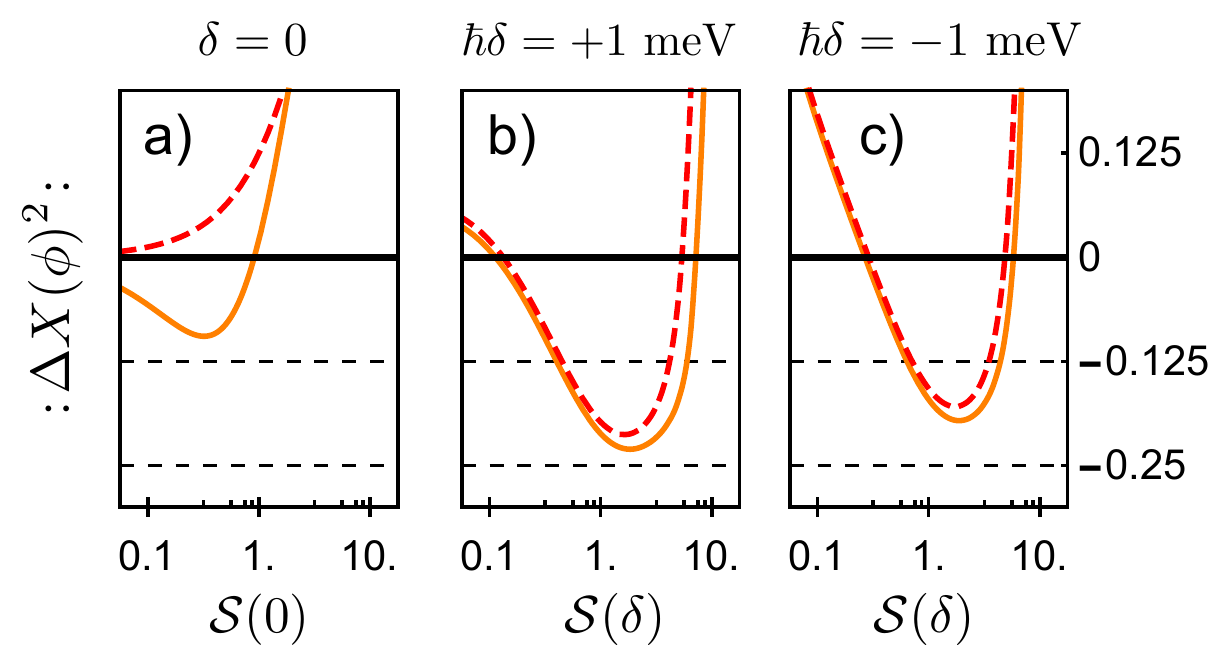}
		\caption{Normalised quadrature variance calculated using our full phonon model without (orange, solid curves), and 
			with (red, dashed curves) additional non-phonon induced pure-dephasing. This dephasing results in no squeezing 
			below saturation on resonance, but the squeezing in the phonon-enhanced regime remains.}
		\label{Dephasing}
	\end{center}
\end{figure*}

Our model described above includes spontaneous emission processes 
and effects caused by coupling to longitudinal acoustic phonons, which can 
be of both dephasing and dissipation in nature. The phonon enhanced coherent scattering 
and associated squeezing processes described in the main text take place in a regime 
where the phonon effects included in the phonon dissipator $K_{\mathrm{ph}}$ dominate. 
As such, the emergence of the phonon enhanced regime is robust against other 
(non-phonon induced) dephasing processes that are weak or moderate compared 
to the spontaneous emission rate. These additional dephasing processes, may be 
caused, for example, by charge fluctuations.

To demonstrate the robust nature of the phonon enhanced squeezing regime, 
in Supplementary Figure~({\ref{Dephasing}}) we show a version of Figure~(2) of the main text, 
which plots the normally ordered quadrature variance as a function of 
driving strength. The solid orange line is our theory including phonons and spontaneous emission but without additional pure-dephasing 
(identical to the solid orange curve in the main text), and the dashed red curve shows the effect of also including pure-dephasing, 
which we achieve by adding a term 
$\gamma(\sigma^{\dagger}\sigma\rho_\mathrm{V} \sigma^{\dagger}\sigma-(1/2)\{\sigma^{\dagger}\sigma,\rho_\mathrm{V}\})$ 
to the Schr\"{o}dinger picture master equation, 
with the dephasing rate $\gamma$ here  
equal to the spontaneous emission rate $\Gamma$. Interestingly, we see 
that this amount of pure-dephasing is sufficient to completely eliminate 
the squeezing below saturation that occurs in the absence of phonons, 
but the squeezing remains in the above saturation regime, with a magnitude nearly equal to that in its absence.

\section*{Supplementary Note 5 - Emitted Field Wigner Functions}

\begin{figure*}
	\begin{center}
		\includegraphics[width=0.95\textwidth]{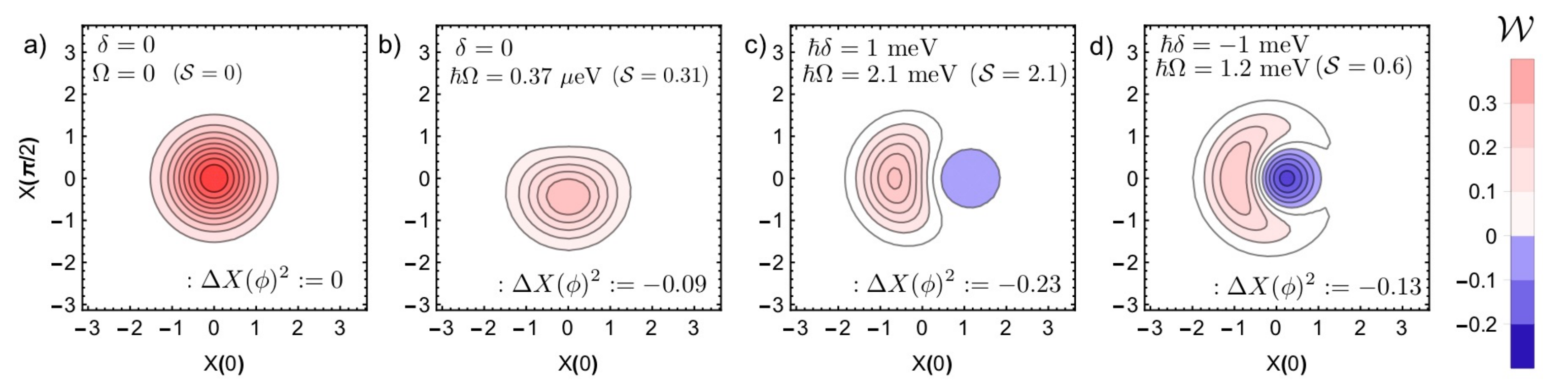}
		\caption{Wigner functions for the vacuum state a), and three squeezed states generated in resonance fluorescence; 
			on resonance and with weak driving b), and off resonant with strong driving c) and d). In cases 
			c) and d) the squeezing occurs due to thermalisation, and can attain levels greater than in case b). Parameters 
			as in all figures in the main text.
		}
		\label{Wigners}
	\end{center}
\end{figure*}

In order to elucidate the nature of the squeezed states of light produced, we can consider the emitted field Wigner function  
defined as $\mathcal{W}(x,p) = \pi^{-1}\int_{-\infty}^\infty \bra{x+y}\rho_{\mathrm{em}}\ket{x-y}\exp[-2i p y]dy$, 
where $\rho_{\mathrm{em}}$ is the state of the field.  
Following Ref.~\cite{Schulte2015} we use the correspondence between the field operators and QD operators 
to associate 
the QD excited state $\ket{e}$ with the first field Fock state $\ket{1}$ and the QD ground state $\ket{g}$ with the field vacuum $\ket{0}$, 
such that $\rho_{\mathrm{em}} = \sum_{n,m = 0,1}(\rho_\mathrm{V})_{nm}\ket{n}\bra{m}$. 
In doing so, we expect $\mathcal{W}(x,p)$ to provide a qualitative representation of the  
electromagnetic field Wigner function for a class of measurements~\cite{PhysRevLett.121.263603}.

In Supplementary Figure~\ref{Wigners} we show Wigner functions for the vacuum a), the squeezed state 
generated in the weak resonant excitation regime below saturation b), and the squeezed states generated only in the presence of phonons 
above saturation for off-resonant driving 
c) and d). Interestingly, in the resonant case b), although the state generated 
is strictly speaking non-Gaussian, its Wigner function is nevertheless positive everywhere, and is not significantly 
dissimilar from a truly Gaussian displaced squeezed vacuum state~\cite{walls2008quantum}. 
However, in cases c) and d) the non-Gaussian nature 
of the field is quite apparent, with the Wigner functions taking on substantial negative values associated with non-classicality. 
Thus, in this regime a highly non-classical, quadrature squeezed and antibunched state 
is produced.

\providecommand{\noopsort}[1]{}\providecommand{\singleletter}[1]{#1}%


\end{document}